\documentstyle[11pt,epsfig]{article}
[1999/03/29 PSNFSS v.7.2
Times font as default roman
: S Rahtz]

\begin{document}
\renewcommand{\thefootnote}{\fnsymbol{footnote}}
\sloppy
\newcommand{\rp}{\right)}
\newcommand{\lp}{\left(}
\newcommand \be  {\begin{equation}}
\newcommand \bea {\begin{eqnarray}}
\newcommand \ee  {\end{equation}}
\newcommand \eea {\end{eqnarray}}

\title{Mechanism for Powerlaws without Self-Organization}
\thispagestyle{empty}

\author{Didier Sornette$^{1,2}$ \\
$^1$ Institute of Geophysics and
Planetary Physics\\ and 
Department of Earth and Space Science\\
 University of California, Los Angeles, California 90095\\
$^3$ Laboratoire de Physique de la Mati\`{e}re Condens\'{e}e\\ CNRS UMR6622 and
Universit\'{e} de Nice-Sophia Antipolis\\ B.P. 71, Parc
Valrose, 06108 Nice Cedex 2, France}

\date{}

\maketitle

email: sornette@moho.ess.ucla.edu

\vskip 1cm

Abstract: A recent claim has been made that there must be a self-regulation 
in the waiting times to see hospital consultants  on the ground
that the {\it relative} changes in the size of waiting lists follow a power law
\cite{Smethurst}. In agreement with simulations of Frecketon and 
Sutherland, we explain the general non-self-regulating 
mechanism underlying this
result and derive the exponent value $-2$ exactly. In addition, we provide
links with related phenomena encountered in many other fields.


\vskip 1cm

\pagenumbering{arabic}

Power law probability distribution (pdf) functions of event or object sizes seem to 
be an ubiquitous statistical features of
natural and social systems. It has repeatedly argued that such an observation
calls for an underlying self-organizing mechanism. In the last
decade, such claims have been made for instance for earthquakes, weather and climate changes,
solar flares, the fossil record, and many other systems, 
to invoke the relevance of self-organized criticality
\cite{Bakhownature}. 
This claim is often unwarranted as there are many non-self-organizing
mechanisms producing power law distributions \cite{sweeping,mybook}.
Recently, a similar claim of the existence of self-regulation
in the waiting time to see hospital consultants has been made on the ground
that the {\it relative} changes in the size of waiting lists follow a power law
\cite{Smethurst}. Using numerical simulations,
Frecketon and Sutherland \cite{Frecketon} show however that
such a power law for {\it relative} changes can be reproduced by 
the simple non-self-regulating random walk null-hypothesis, which obtains
a power law exponent (slope of the log-log plot of the non-cumulative distribution)
indistinguishable from $-2$. 

Here, we explain the general non-self-regulating mechanism underlying this
result and derive the exponent value $-2$ exactly. This mechanism is nothing but 
the effect of the
conservation of probability $P_x(x)~dx = P_y(y)~dy$ on the 
probability distribution function (pdf) $P_x(x)$ 
of a change of a variable $x$ to another $y=f(x)$. In other words,
the objective estimation of the probability of an event is invariant under a change of
mathematical description of this event. If the transformation is an inversion
$y = x^{-1/\alpha}$ with $\alpha>0$, then
\be
P_y(y) = \alpha{P_x(x(y)) \over y^{1+\alpha}}~.  \label{ggnwfqw}
\ee
Suppose that $P_x(x)$ goes to a constant for $x \to 0$, 
then the distribution of $y$ for
large $y$ is a power law with a tail exponent $1+ \alpha$.
The uniform fluctuations of $x$ close to zero lead to scale-free and arbitrarily large
fluctuations of its inverse power $y$. The power law form is kept obviously 
(with a modification of the value of the exponent) if $P_x(x)$ itself
goes to zero or diverges close to the origin as a power law. 

This mechanism operates in many situations. Let us illustrate it in details
in a simple well-controlled situation to demonstrate that it also applies
to rather complex systems. Consider the Ising model of interacting spins
with ferromagnetic interactions which tend to align the spins while thermal 
fluctuations tend to randomize them. The Ising model is a paradigm of the 
fight between order and disorder leading to a
spontaneous organization by collective behavior and has been used to 
describe both physical and social phenomena. Here, we recall that the
distribution of the {\it relative} changes $\Delta M/M$ of magnetization $M$ (total normalized
sum of the spin values) is a power law pdf, but this power law owes nothing
to organization and everything to the ``change-of-variable'' mechanism with 
an inversion.
Right at the critical temperature $T = T_c$, the pdf $P(M)$
of magnetization has a very ``thin'' exponential tail
$P(M) \propto M^{(\delta-1)/2} \exp \{ -{\rm const} ~ M^{\delta+1} \}$,
with $\delta+1 \approx 5.8$ for the 3D Ising universality class
\cite{Bruce95}.  
The distribution of changes $\Delta M$ of the magnetization $M$ under a fixed
number of Monte Carlo steps per site
is a Gaussian of standard deviation $W$ at $T = T_c$ \cite{Binderisingpow,Staufferpowising}.
However, the distribution of the relative changes $\Delta M/M$ is 
a power law with exponent $-2$ \cite{NaeemJanetal}! How does it come about?

Large values of $\Delta M/M$ come from
the limit $M \rightarrow 0$ rather than $\Delta M \rightarrow \infty$. The 
probability $P(M)$ goes to a constant for $M \rightarrow 0$, while 
$\Delta M$ can be approximated by the width $W$ of the Gaussian; thus:
$P(X)dX = P(\Delta M/M) d(\Delta M/M) = P(M) dM \approx {\rm const}~ dM$ 
and $P(X) = P(M)/(dX/dM) = {\rm const}~/(d(W/M)/dM) \propto 1/X^{2}$,
in agreement with Jan et al. \cite{NaeemJanetal}.
The $1/X^2$ power law is in fact not restricted to the
critical point and is very general since it results simply from the inversion mechanism
$y = x^{-1/\alpha}$ defined above with $\alpha =1$. A similar argument
has been proposed in \cite{NaeemJanetal} in this context.

The power law pdf for the random walk
null-hypothesis studied by Frecketon and Sutherland \cite{Frecketon} results from
the choice of the variable which is the
{\it relative} change in numbers $(N_{t+1}-N_t)/N_t$. According to the 
``change-of-variable'' mechanism with an inversion, the power law tail
is controlled by excursions of the denominator towards small values. Since again
$\alpha=1$, this predicts the value $-2$ for the exponent in agreement with the 
numerical simulations\cite{Frecketon}. It is tempting to conclude that the
same general mechanism operates for the results of Smethurst and Williams \cite{Smethurst}.

The same ``change-of-variable'' mechanism with an inversion applies in many other situations
to create power law pdf's (see \cite{mybook} for details and references):
\begin{itemize}
\item In continuous percolation, 
the pdf of transport coefficients such as conductance,
permeability and rupture thresholds and of necks between random holes or random cracks
\cite{Halperinfeng1,Halperinfeng2,Sorperco};

\item The pdf of fluid velocities due to the presence of vortices in hydrodynamics
\cite{Weiss};

\item The Holtsmark's distribution
of gravitional forces created by a random distribution of stars in an infinite 
universe (see \cite{Feller} and chapter 17 of \cite{mybook}
and references therein);

\item This result for the Holtsmark's distribution applies to other fields
with a suitable modification of the exponent, such as electric,
elastic or hydrodynamics, with a singular power law dependence of the force as a function of the
distance to the source (see chapter 17 of \cite{mybook} and references therein).

\item The statistics 
\be
T\equiv {\sqrt{n} (\bar{x} - \langle x \rangle) \over 
[(n-1)^{-1}\sum_{j=1}^n (x_j-\bar{x})^2]^{1/2}}
\ee
 occurs 
in the construction of tests and confidence intervals of the random variable and
is described by Student's distribution, whose power law tail results from the 
same ``change-of-variable'' mechanism with an inversion. Here the inversion refers
to the fact that the estimated standard deviation in the denominator of $T$ can
approach zero leading again to large fluctuations of $T$. For a demonstration, 
see chapter 14 of \cite{mybook}
and references therein.
\item Berry's ``battles of catastrophes'' \cite{berrypdfcata};

\item $-7/2$ power law pdf of density in the 
Burgers/adhesion model of the universe \cite{Frischbecetal}.

\end{itemize}

In sum, this communication has explained one among many other mechanisms
for power law pdf's
(see chapter 14 of \cite{mybook} for a general review) that are
not based on self-regulation or self-organization.
By this understanding of the main possible mechanisms, and complementing this
toolbox with mechanisms
relying on collective effects, such as in percolation, criticality and 
self-organized criticality, it should become possible 
to identify which one is the most relevant to a given problem.

\end{document}